\documentclass[traditabstract]{aa}

\usepackage{graphicx}
\usepackage{multirow}
\usepackage{natbib}
\usepackage{adjustbox}
\usepackage{txfonts}
\usepackage{lipsum} 
\usepackage{mathrsfs}
\usepackage{breqn}
\usepackage[colorlinks = true,linkcolor = blue,urlcolor = blue,citecolor = blue]{hyperref}
\usepackage{xcolor}
\definecolor{yellowgray}{rgb}{0.90, 0.90, 0.2}
\definecolor{bluegray}{rgb}{0.20, 0.60, 0.80}
\definecolor{palered}{rgb}{0.99, 0.40, 0.5}
\definecolor{darkgray}{rgb}{0.35, 0.35, 0.35}
\definecolor{darkgrayb}{rgb}{0.75, 0.75, 0.75}
\definecolor{palegray}{rgb}{0.96, 0.96, 0.96}

\begin{document} 

\title{Multi-wavelength spectropolarimetric observations \\ of AR13724 performed by GRIS}
\titlerunning{Multi-wavelength spectropolarimetric observations with GRIS}

\author{C. Quintero Noda\inst{1,2} \and
        J. C. Trelles Arjona\inst{1,2} \and
        T. del Pino Alemán\inst{1,2} \and
        C. Méndez Lápido\inst{2} \and        
        M. J. Martínez González\inst{1,2} \and
        T. Felipe\inst{1,2} \and
        S. Regalado Olivares\inst{1} \and
        P. Gómez González\inst{1} \and
        J. Bienes\inst{1} \and
        J. Quintero Nehrkorn\inst{1} \and
        A. Matta-Gómez\inst{1} \and
        M. Barreto\inst{1} \and
        M. Collados\inst{1,2}
       }
\institute{Instituto de Astrof\'isica de Canarias, E-38205, La Laguna, Tenerife, Spain   \\
           \email{carlos.quintero@iac.es}    
           \and  Departamento de Astrof\'isica, Univ. de La Laguna, La Laguna, Tenerife, E-38200, Spain}
\date{Received 13 December 2024 ; accepted 17 April 2025}


\abstract{Multi-line spectropolarimetric observations allow for the simultaneous inference of the magnetic field at different layers of the solar atmosphere and provide insight into how these layers are magnetically coupled. The new upgrade of the Gregor Infrared Spectrograph (GRIS) instrument offers such a possibility, allowing for the simultaneous observation of the \ion{Ca}{ii} line at 8542~\AA, the \ion{Si}{i} line at 10827~\AA, and the \ion{He}{i} triplet at 10830~\AA \ in addition to some additional weaker spectral lines that can probe deeper in the photosphere. Because these spectral lines are sensitive to the plasma properties at different regions of the solar atmosphere, their combined analysis can help understand the stratification of its thermal and magnetic properties from the photosphere to the chromosphere. This work showcases recent observations of the upgraded GRIS at the active region AR13724, which shows the instrument's potential for unravelling the most minute details of solar phenomena. In particular, we analyse the spatial distribution of the polarisation signals as well as the distribution of Stokes profiles for different regimes of the magnetic field strength. We also conduct a preliminary data analysis using relatively simple and approximate methods.}


\maketitle

\section{Introduction}

Sunspots are the most prominent solar feature on the visible solar surface and can be observed even by the naked eye. They are magnetic structures where the efficiency of convection is reduced, leading to their lower temperature with respect to the surrounding atmosphere and, in turn, to their appearance as dark features in the photosphere. Generally, sunspots show a dark central region, the umbra, and a less dark surrounding ring, the penumbra. It is in the umbra where the strongest magnetic fields can be found on the solar surface, and these fields are usually oriented close to the local vertical. The magnetic field is typically weaker and more inclined as we move towards the penumbra (for further details, see, for example, the book by \citealt{Stix1989} and the reviews by \citealt{2003A&ARv..11..153S} and \citealt{2011LRSP....8....4B}, and references therein).

In order to uncover the 3D structure of the magnetic field vector in a sunspot, observations of several spectral lines sensitive to different layers of the solar atmosphere are necessary. Multi-channel spectropolarimetry allows for the simultaneous observation of several spectral lines and is thus an invaluable tool for magnetic field diagnostics. For instance, the Daniel K. Inouye Solar Telescope \citep{Rimmele2020} and the European Solar Telescope \citep{QuinteroNoda2022EST} have been designed to allow for these multi-channel observations. This can be achieved by simultaneous observations with multiple instruments specialised in different spectral ranges or by building several spectral channels within a single instrument. 

Two main types of instruments are traditionally used for spectro-polarimetric observations of the Sun. Typically based on Fabry-Pérot interferometry, imaging instruments observe a 2D field-of-view (FOV) within a very narrow and tunable wavelength band. They can provide a very high spatial resolution in combination with adaptive optics and image restoration techniques \citep[e.g., Multi-Object Multi-Frame Blind-Deconvolution,][]{Lofdahl2002,vanNoort2005} and hence have been extensively used by solar observatories. For instance, at ground-based telescopes, examples of this type of instruments are the Telecentric Etalon SOlar Spectrometer \citep[][]{Kenticher1998},  the Interferometric BIdimensional Spectropolarimeter \citep[][]{Cavallini2006}, the CRisp Imaging SpectroPolarimeter \citep[][]{Scharmer2008}, the Gregor Fabry-Pérot Interferometer \citep[][]{2012AN....333..880P}, the CHROMospheric Imaging Spectrometer \citep[][]{Scharmer2017}, and the Visible Tunable Filter \citep[][]{Schmidt2014}. In balloon missions like Sunrise \citep{Solanki2010}, there is the Imaging Magnetograph eXperiment \citep[][]{MartinezPillet2011} and the Tunable Magnetograph \citep{2025arXiv250208268D} while, in space, there is, for instance, the Polarimetric and Helioseismic Imager \citep[][]{2020A&A...642A..11S} onboard Solar Orbiter. These instruments usually scan one or multiple spectral lines sequentially, recording the solar information at specific spectral positions (usually between 5-15 wavelength points) with a total cadence of 10-20 s per spectral line when performing spectropolarimetric measurements. Their main drawback is the lack of simultaneity of the spectral profiles, as different wavelengths are sampled one after the other. Moreover, the cadence must be balanced with the density of the spectral sampling and the number of spectral lines observed.

Slit spectro-polarimeters observe a spectral region along a line on the plane of the sky (e.g., the Gregor Infrared Spectrograph, GRIS, \citealt{Collados2012}, or the Visible Spectro-Polarimeter, \citealt{ViSP}). They can provide spectropolarimetric observations, and by moving the slit, they can scan a 2D FOV while they keep the spectral integrity (i.e., the spectral information is recorded in the sensor at the same time) of profiles. However, the size of the FOV must be balanced with the cadence and degree of spatial simultaneity. In that sense, as the spatial resolution becomes higher thanks to the advent of 4-m class telescopes like DKIST and EST, the time needed to cover a reasonably large spatial domain becomes unattainable and a new type of instrument is needed. The more recently developed Integral Field Spectropolarimeters (IFS) are three-dimensional spectrographs combining the 2D capabilities of imaging instruments and the spectral capabilities of slit spectrographs \citep[see, e.g.,][]{2013JAI.....250007C,2019OptEn..58h2417I}. As a result, they provide spectra that are simultaneous in both spatial and spectral domains. Also, as IFS can perform observations over a two-dimensional field-of-view strictly simultaneously, they can take advantage of image restoration techniques to improve image quality \citep[][]{2025A&A...696A...3L}. Another benefit of scanning a 2D FOV, albeit small, is that IFS instruments can compensate for atmospheric refraction as imaging instruments traditionally do. Their main drawback is that IFS cover the same 3D information as their long-slit counterpart. This means that for the same sensor, the elongated and narrow FOV of the long slit becomes a much smaller (compared to the long side of the slit) rectangular FOV in the case of the IFS. Several types of IFS systems have been successfully tested in solar physics in recent years, such as a subtractive double pass \citep[e.g., ][]{2018SoPh..293...36B,2021SoPh..296...30M}; image slicers; microlens arrays; and optical fibres. The micro-lens prototype has been tested at the SST with the microlensed hyperspectral imager instrument \citep[][]{2022A&A...668A.149V} and the Helium Spectropolarimeter. In the case of image slicers, one unit is installed at the Gregor telescope \citep{Schmidt2012,Kleint2020} as an upgrade of GRIS, and it was the first solar IFS system offered openly to the community since 2019 \citep{DominguezTagle2022}. Also, a second unit has been recently installed at the Diffraction Limited Near Infrared Spectropolarimeter \citep{2022SoPh..297..137J} as an upgrade over the original optical fibres-based IFS, improving the capabilities of the instrument \citep[][]{2024SPIE13096E..26A}.

Recently, GRIS has been upgraded to allow for the simultaneous observation of multiple spectral ranges. The version of the instrument offered in 2024 consists of two different cameras installed in each of the two optical arms \citep[see][for the optical design]{RegaladoOlivares2024GRIS}. Later on, a third channel will be commissioned as well. One of the possible configurations for the upgraded GRIS instrument is to simultaneously observe the \ion{Ca}{ii} line at 8542~\AA\ (channel 2) and the spectral region, including the \ion{Si}{i} line at 10827~\AA\ and the \ion{He}{i} triplet at 10830~\AA \ (channel 1). The \ion{Si}{i} line is sensitive in the upper photosphere, the \ion{Ca}{ii} line is sensitive to the lower and middle chromosphere \citep[see, e.g.,][]{QuinteroNoda2016}, and the \ion{He}{i} triplet is useful for the study of filaments and prominences \citep[e.g.,][]{2012A&A...539A.131K,2014A&A...566A..46O,2015ApJ...802....3M}, spicules \citep[e.g.,][]{1969SoPh....7..351B,2015ApJ...803L..18O}, and the on-disk chromosphere \citep[e.g.,][]{2003Natur.425..692S,2009ApJ...692.1211C,2013ApJ...768..111S}. Consequently, combining these spectral lines provides coverage over an extensive range of heights in the solar atmosphere.

The upgraded GRIS started its commissioning in early 2024 (albeit with only the long-slit configuration available), and so far, it has been delivering high spatial and spectral resolution observations with high polarimetric sensitivity \citep{QuinteroNoda2024First}. Moreover, the instrument will offer the option of observing using the integral field unit (as an IFS), hence, the most capable observation mode GRIS can offer from early 2025 on.

\section{Data description}\label{Method}

\begin{figure}
	\begin{center} 
		\includegraphics[width=.48\textwidth]{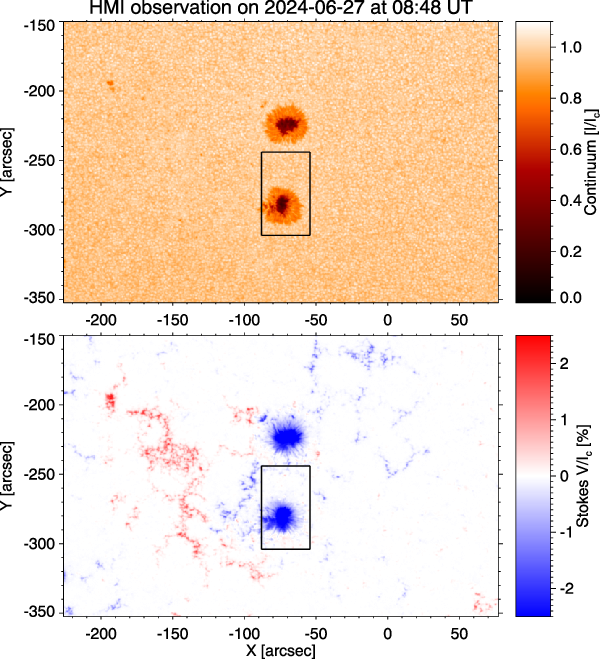}
		\caption{Continuum intensity map near the \ion{Fe}{i} line at 6173~\AA\ (top panel) and Stokes $V$ polarity map for the same spectral line (bottom panel) acquired by the HMI/SDO on 2024 June 27 at 08:48~UT. Both quantities are normalised to the averaged continuum intensity over a quiet Sun area at the disc centre, $I_{\rm c}$. The black square delimits the field of view of our observations with the GRIS instrument.}
		\label{ContextHMI}
	\end{center}
\end{figure}

\begin{figure}
	\begin{center} 
	 \includegraphics[width=.48\textwidth]{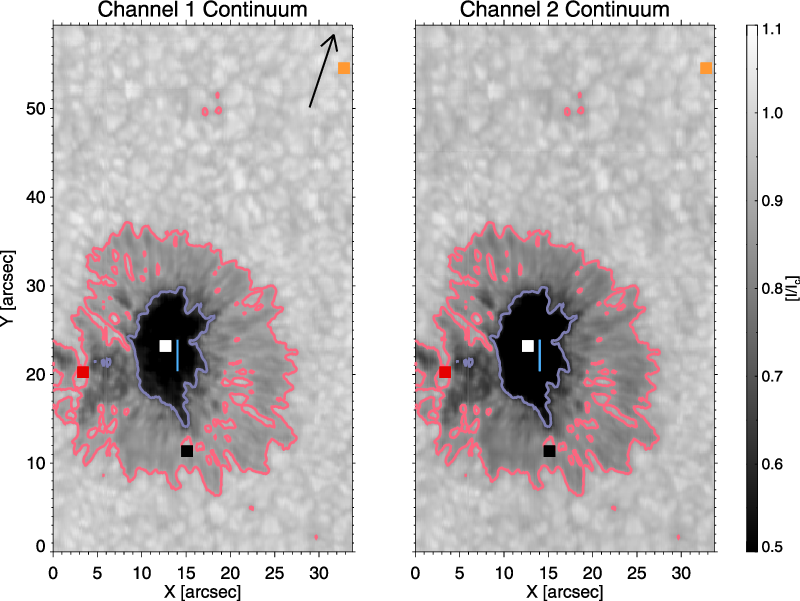}
		\caption{Spatial distribution of the continuum intensity normalised to the averaged continuum intensity over a quiet Sun area located at the top of the observed field of view, $I_{\rm c}$. Panels represent the recorded intensity in channels 1 (left panel) and 2 (right panel) of GRIS. The long-slit scan was performed from the left to right side of the image. The vertical axis corresponds to the north-south direction of the Sun. The coloured symbols highlight regions of interest studied in Section~\ref{Results}. At the same time, contours aim to tentatively underline the location of the umbra (defined as regions with $I_c \sim 0.6$) and the penumbra (areas where $I_c \sim 0.8$). The arrow in the left panel points towards the disc centre.}
		\label{Context}
	\end{center}
\end{figure}

\begin{figure}
	\begin{center} 
		\includegraphics[width=.48\textwidth]{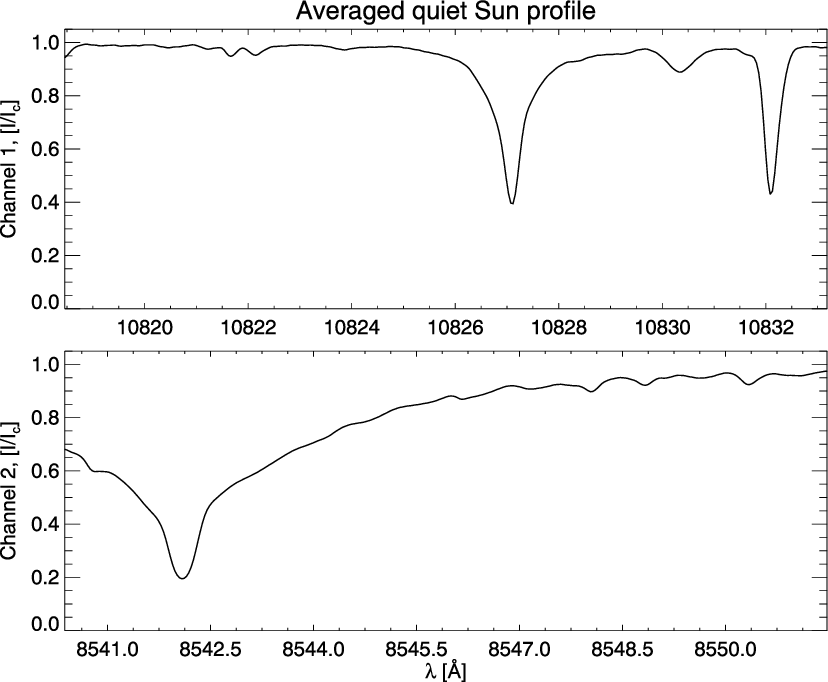}
		\caption{Average quiet Sun intensity profile, normalised to the average quiet Sun continuum intensity, in channels 1 (top panel) and 2 (bottom panel). The average is taken from the quieter area at the top of the observed FOV (see Figures~\ref{ContextHMI} and \ref{Context}).}
		\label{Wave_range}
	\end{center}
\end{figure}

We observed AR13724 with GRIS on 2024 June 27, at 08:50~UT. The active region was close to the disc centre at around $(-76, -240)$~arcsec, at a heliocentric angle of $\mu=0.96$. The observed sunspot was relatively close to another spot with the same polarity as seen in Fig.~\ref{ContextHMI} that displays the observation by the Helioseismic Magnetograph Imager \citep[HMI,][]{2012SoPh..275..207S} onboard the Solar Dynamics Observatory \citep[SDO,][]{2012SoPh..275....3P}. Both sunspots exhibit a rounded shape and a clear penumbra. There were no nearby sunspots with opposite polarity, and their companion polarity is thus the active plage to the left of both sunspots. Our observation includes only the sunspot at the bottom (see black square in Fig.~\ref{ContextHMI} delimiting our FOV).

The observation consisted of a long-slit scan with 250 positions with a step of 0\farcs135. The slit is 60\farcs48 long, with a sampling of 0.135~arcsec/pix over 448 pixels. The FOV covered by the scan is thus roughly $34\times60$~arcsec$^2$ (see Fig.~\ref{Context}). The spatial scale is not strictly the same in the two spectral channels, being about 0.130~arcsec/pix for channel 2. We thus re-scaled the observations from the second channel to match the spatial scale of the first channel.

Figure~\ref{Wave_range} shows the spectral range of the two optical channels of the GRIS instrument. The new channel 2 is centred at 8547~\AA\ and includes most of the \ion{Ca}{ii} line at 8542~\AA. Channel 1 is centred at 10825.75~\AA\ and covers the \ion{Si}{i} line at 10827~\AA\ and the \ion{He}{i} triplet at 10830~\AA. The spectral coverage is about 15 and 11~\AA\ for channels 1 and 2, respectively. The spectral sampling is about 14.7 and 11.1~m\AA\ for channels 1 and 2, respectively. To improve the signal-to-noise ratio, we binned the profiles along the spectral domain every three pixels, resulting in a spectral sampling of about 44.1 and 33.3~m\AA\ for channels 1 and 2, respectively.

We performed a modulation with four polarisation stages \citep{Collados2012,QuinteroNoda2024PMU} with an integration of 100~ms each, resulting in 400~ms for each modulation cycle. For each slit position, we acquired ten modulation cycles (i.e., ten accumulations), so each slit position takes about 4~s. For 250 steps, the total observing time was about 30~min (including the additional time required to move the slit). The reference direction for the linear polarisation corresponds to positive $Q$ in the solar south-north direction. We estimate that the noise at continuum wavelengths is of about $9$, $11$, and $8.5$ in units of $10^{-4}$ the quiet Sun continuum intensity for Stokes $Q$, $U$, and $V$, respectively, for channel 1, and of about $13$, $15$, and $15$ for channel 2, in the same units.

We note that the observation does not reach the continuum in channel 2, which is beyond the 8551.45~\AA\ limit. To normalise the profiles in channel 2 to the average quiet Sun continuum intensity, we assumed for this last wavelength the same relative intensity with respect to the continuum as in the BAse de données Solaire Sol (BASS2000) solar spectrum \citep{Delbouille1973}. In addition, the set-up for the second spectral channel may look suboptimal, with the \ion{Ca}{ii} line centre significantly shifted to the left side of the spectral range. The reason for this is partly optomechanical. The current configuration is part of the preparation for a possible three-channel configuration (the third channel would be at 7700~\AA, covering the \ion{K}{i} D$_1$ line, combined with the configuration shown in this work). The space in the spectrograph and size of the optical components do not allow for better centring of the \ion{Ca}{ii} line when also observing the \ion{He}{i} triplet. Nevertheless, once the third channel is available, it will be possible to use only two of the channels to observe the \ion{Ca}{ii} and \ion{He}{i} lines centred on their respective detectors.

Moreover, because we are observing with the long slit mode of GRIS, we need to consider the impact of the differential atmospheric refraction to discuss the simultaneity of the observations in the two spectral ranges. We have estimated the spatial difference at the date and time of our observation between the two wavelength regions, which is approximately 0.28~arcsec in the case that the shift induced by the atmospheric refraction is strictly perpendicular to the slit. However, when comparing the spatial distribution of intensity signals at the continuum wavelengths of both spectral windows, we obtain the highest correlation when we shift one image with respect to the other 0.135~arcsec horizontally (along the scanning direction). This size corresponds to one slit position, e.g., around 7~s time difference between the two spectral channels. Thus, we consider that we are in a favourable scenario where atmospheric refraction has a low impact on the multi-wavelength configuration used in this work. However, we also want to clarify that a worse scenario can appear when the spectral regions are further apart or the telescope conditions (e.g., a lower elevation) are less favourable. Finally, we also want to note that the GRIS instrument can be configured as an IFS to avoid this limitation. This IFS configuration will be offered starting the first semester of 2025.

\section{Spatial and spectral features}\label{Results}

In this section, we analyse the spatial distribution of the intensity and the polarisation signals for the three spectral lines mentioned in section~\ref{Method}. We also delve into their characteristic Stokes profiles corresponding to different areas of interest in the FOV, namely in the quiet Sun, the penumbra, the umbra, and a light bridge-like area.

\begin{figure*}
	\begin{center} 
		\includegraphics[width=.80\textwidth]{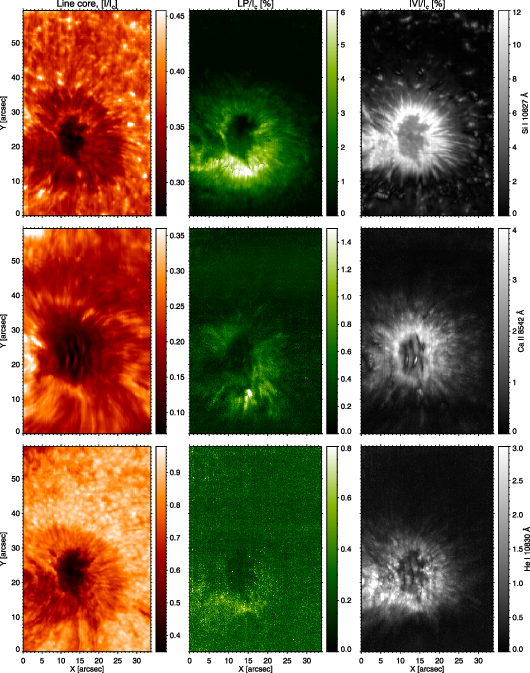}
		\caption{Spatial distribution of the line core intensity (left column), maximum linear polarisation (central column), and maximum circular polarisation in absolute value (right column) for the \ion{Si}{i} line at 10827~\AA\ (top row), the \ion{Ca}{ii} line at 8542~\AA\ (middle row), and the \ion{He}{i} triplet at 10830~\AA \ (bottom row). All quantities are normalised to the average quiet-Sun continuum intensity, $I_{\rm c}$.}
		\label{Spatial_spectra}
	\end{center}
\end{figure*}

\begin{figure*}
	\begin{center} 
		\includegraphics[trim=0 0 0 0,width=18.0cm]{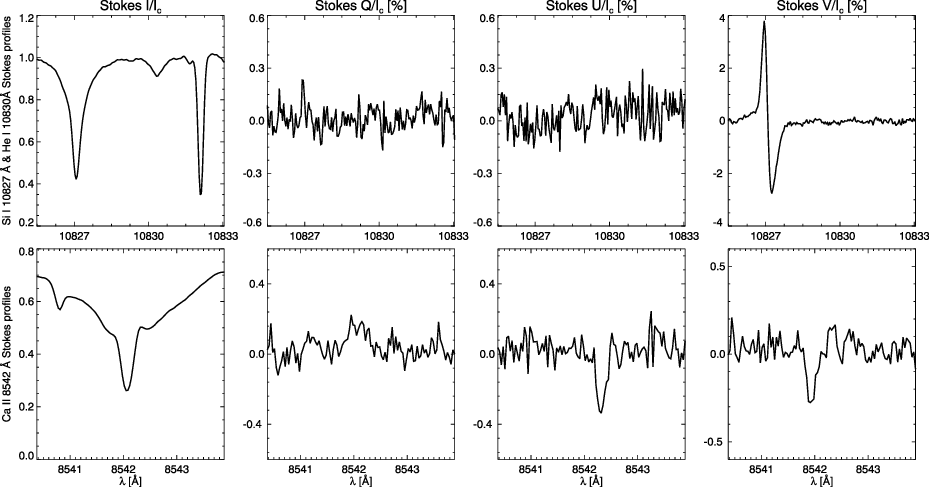}
		\caption{Intensity (first column), Stokes $Q$ (second column), $U$ (third column), and $V$ (fourth column) profiles, normalised to the average quiet-Sun continuum intensity, $I_{\rm c}$. The top row shows the spectra in channel 1, which includes the \ion{Si}{i} line at 10827~\AA\ and the \ion{He}{i} triplet at 10830~\AA. The bottom row shows the spectra in channel 2, which includes most of the \ion{Ca}{ii} line at 8542~\AA. The displayed profiles correspond to the pixel at $(33, 57)$~arcsec, the orange square in Fig.~\ref{Context}, associated with a quiet Sun area with low magnetic activity.}
		\label{Network}
	\end{center}
\end{figure*}

\begin{figure*}
	\begin{center} 
		\includegraphics[trim=0 0 0 0,width=18.0cm]{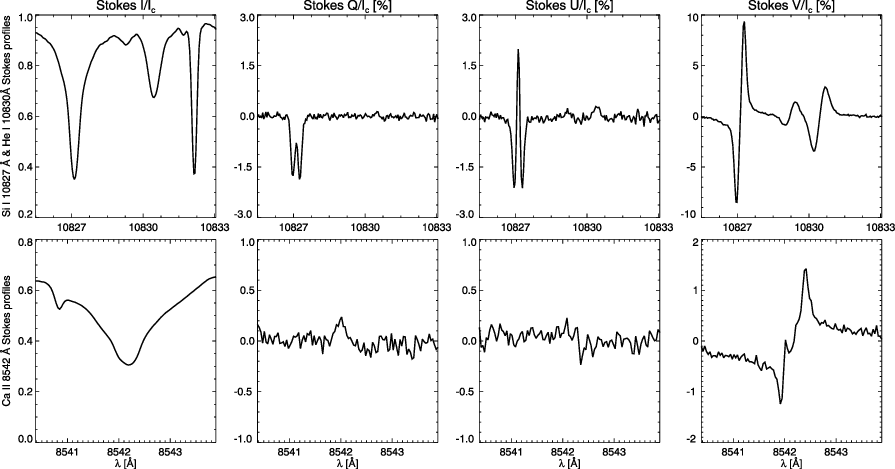}
		\caption{Same as Fig.~\ref{Network}, but for a pixel in the light bridge-like area at $(3.4, 20.3)$~arcsec, the red square in Fig.~\ref{Context}.}
		\label{LB}
	\end{center}
\end{figure*}

\subsection{Spatial distribution}

Figure~\ref{Spatial_spectra} shows the spatial distribution of the line core intensity, the maximum total linear polarisation (calculated as $LP = \sqrt{Q^2+U^2}$), and the maximum circular polarisation in absolute value, for the \ion{Si}{i} 10827~\AA, the \ion{Ca}{ii} 8542~\AA, and the \ion{He}{i} 10830~\AA \ spectral lines. For the \ion{Si}{i} transition (top row), we see the umbra and penumbra surrounded by the reverse granulation pattern. In the \ion{Ca}{ii} spectral line (middle row), the dark umbra seems to fill a larger area when compared to the \ion{Si}{i} map, showing the presence of a thread-like pattern. The thread-like structures in the umbra in the \ion{Ca}{ii} line seem to be an indication of the presence of umbral flashes \citep{1969SoPh....7..351B}. In the case of the helium line, we detect many areas that resemble the spatial distribution of continuum signals (see Figure~\ref{Context}). Still, there are other regions where we detect dark threaded features, such as those in the bottom left part of the FOV.

The maximum linear polarisation map for \ion{Si}{i} presents relatively weak signals in the regions further from the sunspot and the umbra. However, there are relatively intense polarisation signals localised in the penumbra. For \ion{Ca}{ii} and \ion{He}{i}, the linear polarisation signals are generally weak, even in the penumbra, and only some localised regions show signals whose amplitude is above the noise.

The spatial distribution of circular polarisation signals is similar for all the transitions, with significant values in the inner penumbra and localised areas outside the sunspot. Especially for the \ion{Si}{i} spectral line, circular polarisation signals are found at positions of enhanced brightness in the line core intensity. Also, while the circular polarisation signals seem dimmer in the umbra, this is just due to its overall smaller brightness and the fact that the spectra are normalised to the averaged quiet Sun continuum intensity. Moreover, the signature of the mentioned thread-like features or umbral flashes can also be seen in the \ion{Ca}{ii} circular polarisation \citep[see, e.g.,][]{2013A&A...556A.115D} and in the \ion{He}{i} line as well \citep[for example,][]{2006ApJ...640.1153C,2015LRSP...12....6K}. In contrast to the linear polarisation signals, the silicon and calcium lines show strong circular polarisation signals in the upper (northern) part of the umbra, as shown in Figures \ref{Context} and \ref{Spatial_spectra}. This indicates that the magnetic field exhibits different degrees of inclination with respect to the local vertical on each side of the penumbra, as will be seen in section~\ref{SS-Btheta}. 

\subsection{Characteristic profiles}

In this section, we showcase the observational capabilities of the upgraded GRIS instrument by showing and analysing spectro-polarimetric profiles characteristic of the different features found within the FOV of our sunspot observation (see section~\ref{Method}). In particular, we have selected profiles in the quiet Sun, penumbra, umbra, an umbral flash, and a light bridge-like area.

\subsubsection{Quiet Sun}

The analysis of polarisation signals in the quiet Sun \citep[e.g.,][]{2003A&A...408.1115K,2007A&A...469L..39M,2016A&A...596A...6L} is challenging due to the generally small amplitude of the signals \citep[see, for instance, the reviews by][]{2017SSRv..210..109D,2022ARA&A..60..415T}. For the polarimetric noise of our observation, namely $\sim10^{-3}I_{\rm c}$, most of the quiet Sun is at the level or below the noise. However, in those regions with magnetic field concentrations, such as network patches, as in, for instance, $(33, 57)$~arcsec (see the orange square in Fig.~\ref{Context}), the polarisation amplitude is well above the noise level.

Figure~\ref{Network} shows the Stokes parameters at the abovementioned position corresponding to a magnetic network pixel. The intensity in the \ion{Si}{i} spectral line is similar to that of the average quiet Sun in Fig.~\ref{Wave_range}. The linear polarisation signals are weak and close to the noise level. However, two positive lobes of small amplitude can be barely identified in the Stokes $Q$ profile. The circular polarisation shows an antisymmetric signal with a red lobe wider than the blue one, which displays a larger amplitude instead. The intensity profile for \ion{He}{i} is also similar to the average quiet Sun profile in Fig.~\ref{Wave_range}, but it does not exhibit polarisation signals above the noise level. The \ion{Ca}{ii} intensity profile has an apparent asymmetry and enhanced intensity in its red wing. The linear and circular polarisation signals are detectable as they are slightly above the noise level.

\subsubsection{Light bridge-like area}

Figure~\ref{LB} displays the Stokes parameters at $(3.4, 20.3)$~arcsec, the red square in Fig.~\ref{Context}. In this region located to the left of the umbra in the figure, the filamentary pattern of the penumbra is disrupted by a dark feature with bright roundish dots and one elongated bright intrusion, which could be a weak umbra with umbral dots and a light bridge \citep[e.g.,][]{1997ApJ...484..900L}. The polarisation signals in the \ion{Si}{i} spectral line are significant and above the noise level. Interestingly, although linear polarisation profiles show a small degree of asymmetry in amplitude, the Stokes $V$ signals seem uniform in terms of amplitude and area asymmetries. The \ion{He}{i} triplet absorption is stronger than in the quiet region, with signals that are small in linear polarisation but significant in circular polarisation. The \ion{Ca}{ii} intensity profile shows an apparent asymmetry, likely due to line-of-sight velocity gradients in the underlying atmosphere. Unlike the significant circular polarisation signal, the linear polarisation profiles are relatively weak.

\begin{figure*}
	\begin{center} 
		\includegraphics[trim=0 0 0 0,width=18.0cm]{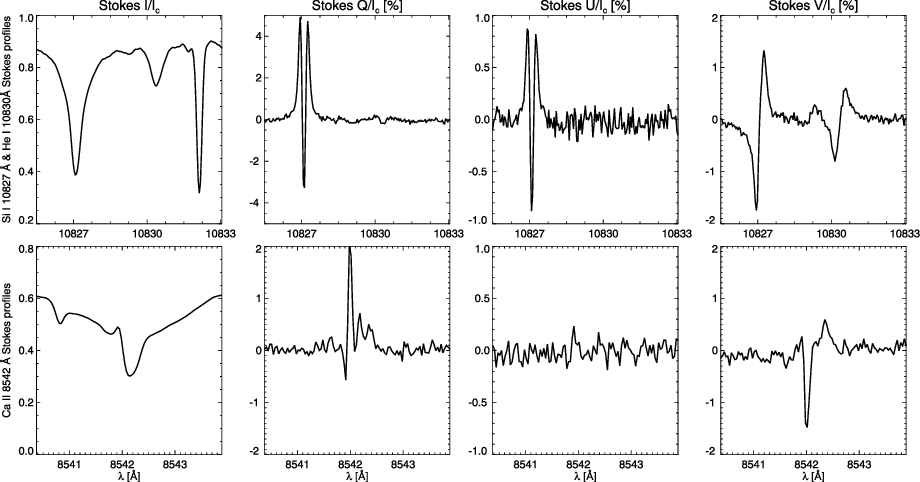}
		\caption{Same as Fig.~\ref{Network}, but for a pixel in the penumbra at $(15.1, 11.3)$~arcsec, the black square in Fig.~\ref{Context}.}
		\label{Pen}
	\end{center}
\end{figure*}

\subsubsection{Penumbra}

In Fig.~\ref{Pen}, we showcase the Stokes profiles at $(15.1, 11.3)$~arcsec, the black square in Fig.~\ref{Context}. This pixel is located in the sunspot penumbra. As expected, the linear polarisation signals are strong in the \ion{Si}{i} and \ion{Ca}{ii} transitions, as the magnetic field is strongly inclined with respect to the local vertical, which is almost parallel to the line of sight (LOS) as the observations are made close to disk centre, in these regions. Circular polarisation is still detected above the noise level, indicating that this dataset may allow us to infer the topology of the magnetic field in sunspots with reasonably high accuracy.

\begin{figure*}
	\begin{center} 
		\includegraphics[trim=0 0 0 0,width=18.0cm]{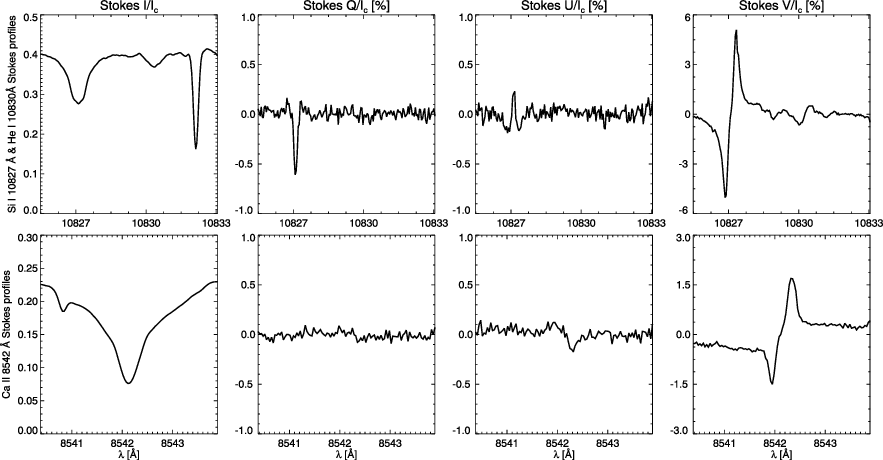}
		\caption{Same as Fig.~\ref{Network}, but for a pixel in the umbra at $(12.7, 23.9)$~arcsec, the white square in Fig.~\ref{Context}.}
		\label{Umbra}
	\end{center}
\end{figure*}

\subsubsection{Umbra and umbral flashes}

\begin{figure*}
	\begin{center} 
		\includegraphics[width=.98\textwidth]{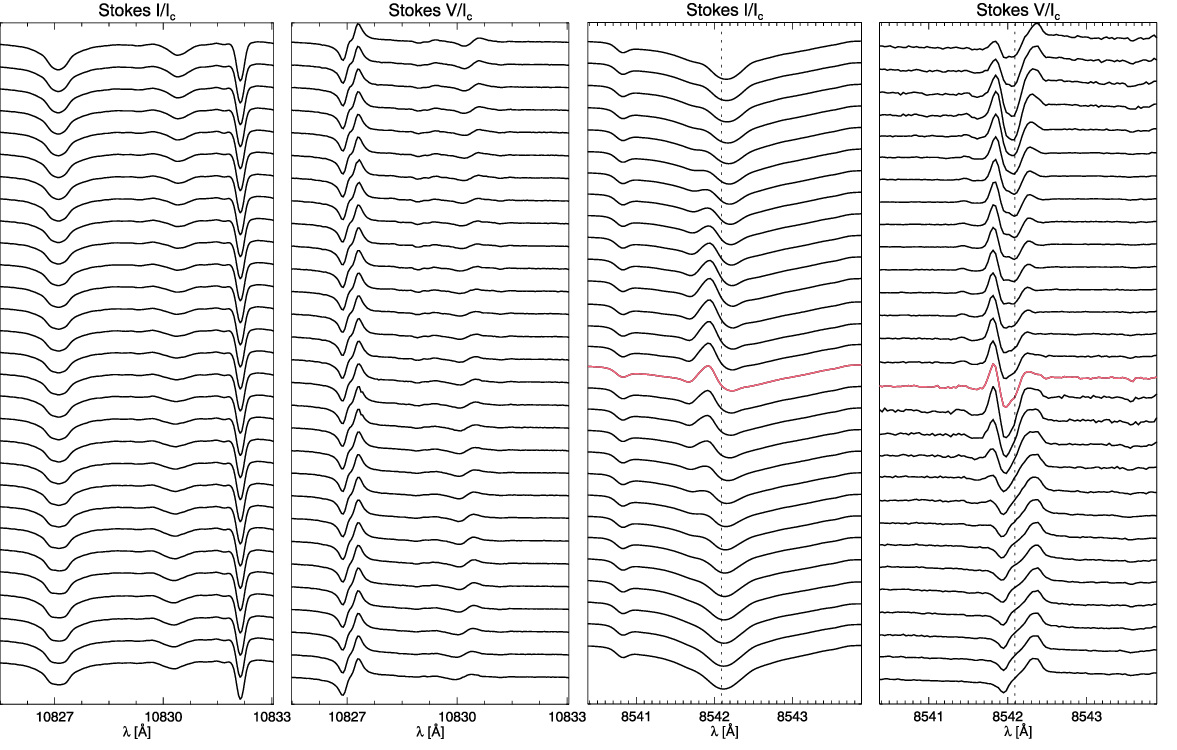}		
		\caption{Intensity (first and third panels from the left) and circular polarisation (second and fourth panels from the left) profiles for channel 1 (first and second panels) and channel 2 (third and fourth panels) for all pixels in a region that covers 4~arcsec (i.e., 29 pixels) along the spectrograph slit indicated with a blue segment in Fig.~\ref{Context}. Red highlights the pixel we study in Figure~\ref{Deriv}.}
		\label{UmbraIV}
	\end{center}
\end{figure*}

\begin{figure}
	\begin{center} 
		\includegraphics[width=.48\textwidth]{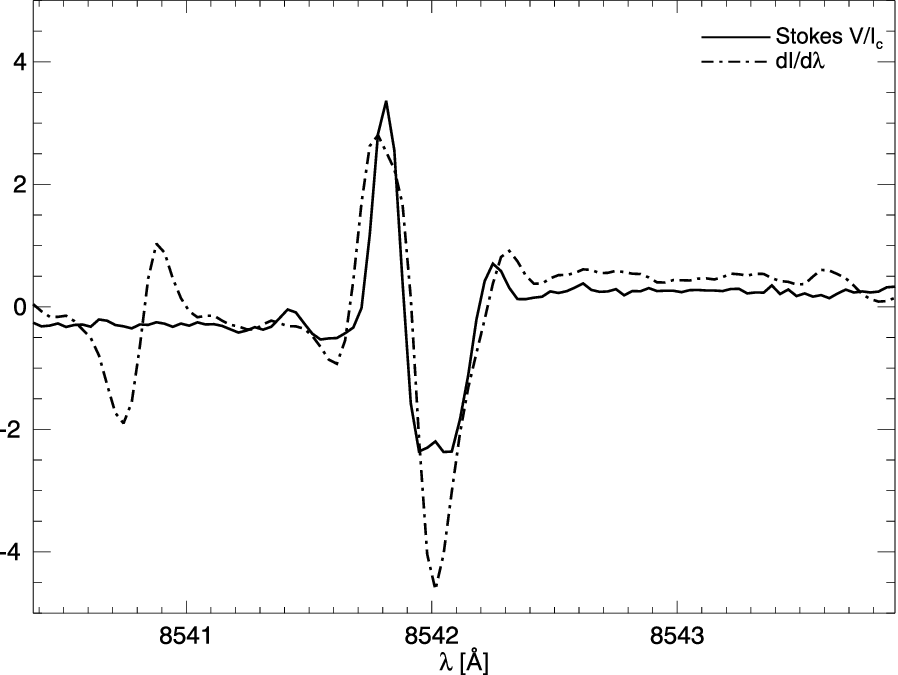}	
		\caption{Comparison between the Stokes $V$ profile (solid) and the derivative of Stokes $I$ with respect to the wavelength. The Stokes profiles are normalised to the quiet Sun averaged continuum intensity, and the representation of Stokes $V$ is in percentage units. At the same time, we scale the derivative with an arbitrary factor to match the Stokes $V$ amplitude. The location of this pixel is highlighted in colour in Figure~\ref{UmbraIV}.}
		\label{Deriv}
	\end{center}
\end{figure}

Figure~\ref{Umbra} shows the Stokes parameters at $(12.7, 23.9)$~arcsec, the white square in Fig.~\ref{Context}. The \ion{Si}{i} intensity profile displays a complex shape, likely due to the significant magnetic field strength. Regarding its polarisation, we observe substantial signals of circular polarisation and weak linear polarisation signals. The presence of the latter indicates that the magnetic field is not fully aligned with the LOS. The absorption in \ion{He}{i} is weaker than in the average quiet Sun (see Fig.~\ref{Wave_range}), and only its circular polarisation is above the noise level. Finally, in the \ion{Ca}{ii} spectral line, we only observe significant circular polarisation as the amplitude of the linear polarisation is below the noise level.

While the umbra is predominantly dark when observed in the line centre of the \ion{Ca}{ii} line, we can see the signature of umbral flashes (see Fig.~\ref{Spatial_spectra}). They are sudden brightenings close to the line core of some chromospheric transitions in sunspot umbrae, produced by acoustic waves travelling upwards from the photosphere and becoming shocks due to the steep fall of the plasma density. Umbral flashes have been thoroughly studied using spectropolarimetric observations with long-slit spectrographs such as the Tenerife Infrared Polarimeter \citep[][]{2007ASPC..368..611C} at the Vaccum Tower Telescope \citep[][]{1998NewAR..42..493V} or imaging instruments such as the CRISP at the SST. As examples, we have the works of \cite[e.g.,][]{2001ApJ...550.1102S,2013A&A...556A.115D,2017ApJ...845..102H,2018A&A...619A..63J} that, except for a few exceptions \citep{2013SoPh..288...73M,2023ApJ...945L..27F}, have mainly employed imaging spectropolarimeters. One of the drawbacks of imaging instruments is the time-dependent acquisition of the wavelengths along the spectral line, which can produce deformations of the Stokes profiles when the solar atmosphere changes at short temporal scales, as in the case of umbral flashes \citep{2018A&A...614A..73F}. In this regard, while the particular observation shown in this work is a slit scan, the analysis of umbral flashes using IFS spectrographs such as GRIS can potentially improve our understanding of this phenomenon by providing better spectral resolution and the simultaneous acquisition of all the wavelengths of the spectral line (better spectral integrity).

Figure \ref{UmbraIV} shows the intensity and circular polarisation profiles for the \ion{Ca}{ii}, \ion{Si}{i} and \ion{He}{i} spectral lines along a region of the spectrograph slit at one step of the scan, indicated by the blue segment in Fig.~\ref{Context}. The linear polarisation profiles are not included since their signal is below the noise level. The \ion{Si}{i} spectral line and \ion{He}{i} triplet do not show any remarkable changes in the selected region of the umbra, something we will study in more detail in the future. In contrast, there is significant variability of the spectra along the slit in the \ion{Ca}{ii} transition. Towards the borders of this region, the profiles show a similar shape as those from the sunspot dark umbra (see Fig.~\ref{Umbra}). However, near the centre of the selected region, we observe a striking intensity enhancement at the blue wing of the \ion{Ca}{ii} line that could be due to the propagation of waves from deeper layers \citep[see, for instance, the series of works that started with][]{2009A&A...507..453B}. The umbral flash also manifests in the circular polarisation signals. They exhibit three-lobed profiles that still seem to correspond approximately to the first derivative of the intensity profile (see Figure~\ref{Deriv}) suggesting the applicability of the weak-field approximation (WFA) for analysing the data (Sect.~\ref{SS-Btheta}).

\subsection{Inferred physical quantities}

In this section, we apply relatively simple diagnostic techniques to obtain a first estimation of relevant physical quantities in the solar atmosphere at the region of formation of the observed spectral lines. In particular, we study the velocity along the LOS and apply the WFA to the Stokes profiles to estimate the magnetic field vector. We plan to use more sophisticated inversion techniques for this dataset and present the results in a following publication.

\begin{figure*}
	\begin{center} 
		\includegraphics[width=.95\textwidth]{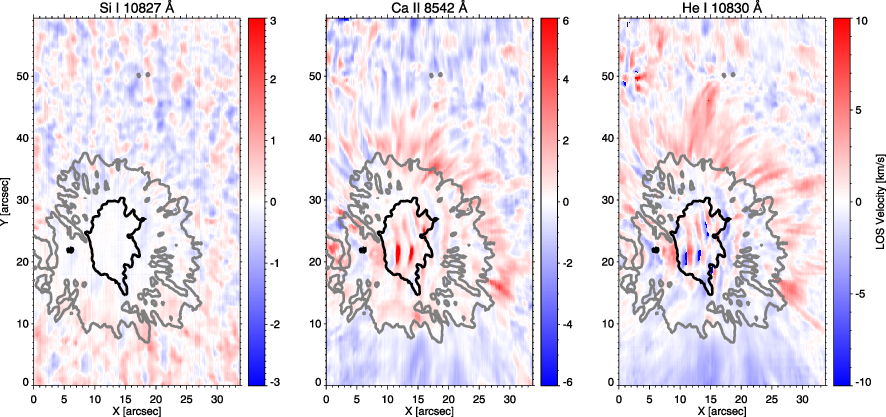}
		\caption{Spatial distribution of the velocity along the LOS inferred from the Doppler shift of the \ion{Si}{i} 10827~\AA \ (left), \ion{Ca}{ii} 8542~\AA \ (middle), and \ion{He}{i} 10830~\AA\ (right) spectral line centres. Contours tentatively define the area corresponding to the umbra and penumbra computed from the continuum intensity maps presented in Figure~\ref{Context}.}
		\label{LOSvel}
	\end{center}
\end{figure*}

\subsubsection{LOS velocities}

We compute the velocity along the LOS from the Doppler shift of the line centre of each spectral line with respect to its theoretical line centre (see Fig.~\ref{LOSvel}). We expect this simple approach to be less accurate when the magnetic field strength is such that the Zeeman splitting becomes comparable with the line width, which may happen, e.g., for the \ion{Si}{i} line in the umbra. The positive values (red colours) correspond to redshifts.

The velocity along the LOS at the region of formation of the \ion{Si}{i} line centre shows a spatial distribution that resembles the granulation pattern. The penumbra and the outer regions of the umbra show small velocities, as expected. The velocity along the LOS at the height of formation of the \ion{Ca}{ii} line centre shows an umbra dominated by the thread-like shape of the umbral flashes. The velocity in the penumbra shows a filamentary pattern similar to the one observed in the line centre intensity (see Fig.~\ref{Spatial_spectra}). In the top region of the FOV, we see elongated features that are larger in area when compared with the velocity map for \ion{Si}{i}. Finally, the spatial distribution of velocities from the \ion{He}{i} triplet resembles that of the calcium transition but with larger values and longer extension. The larger velocity values in the \ion{He}{i} line could be due to the higher height of formation and the presence of waves propagating through the atmosphere of the umbra \citep[e.g.,][]{2006ApJ...640.1153C}.

\subsubsection{Magnetic field vector}\label{SS-Btheta}

The WFA \citep[e.g.,][]{1967AnAp...30..257R} equations for the Stokes parameters are obtained from the radiative transfer equations under the assumption that the Zeeman splitting is much smaller than the typical width of the spectral line. Moreover, under the assumption that several physical quantities are constant along the region of formation of the polarisation profiles, it is possible to write closed equations relating the emergent Stokes parameters and the magnetic field vector. These closed equations are \citep[see, e.g.,][]{2004ASSL..307.....L}

\begin{subequations}
\begin{align}
Q = & - \frac{1}{4}C^2 G_{\rm eff} \lambda_0^4  B_{\perp}^2 \cos 2\phi \frac{\partial^2 I}{\partial \lambda^2}, \label{wfaq}\\
U = & - \frac{1}{4}C^2 G_{\rm eff} \lambda_0^4  B_{\perp}^2 \sin 2\phi \frac{\partial^2 I}{\partial \lambda^2}, \label{wfau}\\
V = & - C g_{\rm eff} \lambda_0^2  B_{\parallel} \frac{\partial I}{\partial \lambda}, \label{wfav}
\end{align}
\end{subequations}
where $B_{\parallel}$ is the magnetic field component parallel to the LOS, $B_{\perp}$ is the module of the magnetic field vector in the plane of the sky, $\phi$ is the azimuth of the magnetic field vector in the plane of the sky, $C=4.67\cdot10^{-13}$~G$^{-1}$\AA$^{-2}$, $\lambda_0$ is the line centre wavelength in \AA, $g_{\rm eff}$ is the effective Land\'e factor, and $G_{\rm eff}$ is a number that plays the role of the effective Land\'e factor for linear polarisation.

Note that Eq.~\eqref{wfav} is only valid if $B_{\parallel}$ is constant across the region of formation of Stokes $V$. The restrictions for Eqs.~\eqref{wfaq}--\eqref{wfau} are significantly more strict. They are valid for the line centre if $B_{\perp}$, $\phi$, and the velocity along the LOS are constant across the region of formation. For these expressions to be valid for the whole spectral line, the absorption profile must have a Gaussian shape, and the line widths must be constant with height \citep[see also][]{2018ApJ...866...89C}. Alternatively, the expressions \citep{2004ASSL..307.....L}
\begin{subequations}
\begin{align}
Q = & \frac{3}{4}C^2 G_{\rm eff} \lambda_0^4  B_{\perp}^2 \cos 2\phi \frac{1}{\lambda - \lambda_0}\frac{\partial I}{\partial \lambda}, \label{wfaq2}\\
U = & \frac{3}{4}C^2 G_{\rm eff} \lambda_0^4  B_{\perp}^2 \sin 2\phi \frac{1}{\lambda - \lambda_0}\frac{\partial I}{\partial \lambda}, \label{wfau2}
\end{align}
\end{subequations}
are valid in the line wings (at $\lambda$ such that $|\lambda - \lambda_0|$ is much larger than the Doppler width) if $B_\perp$, $\phi$, and the velocity along the LOS are constant in the region of formation.

\begin{table*}
\caption{Effective Land\'{e} factor for the linear and circular polarisation.}
\begin{center}
\normalsize
\begin{adjustbox}{width=0.75\textwidth}
  \bgroup
\def\arraystretch{1.25}
\begin{tabular}{lcccccccccccccc}
	\hline
Atom & $\lambda$ [\AA] &	Conf$_1$ & Conf$_2$  & $g_1$ & $g_2$ & $g_{\rm eff}$ & $s$ & $d$ & $G_{\rm eff}$ \\
	\hline
Ca~{\sc ii}  & 8542.09   & ${}^2D_{5/2}$     & ${}^2P^{o}_{3/2}$ &  1.20  &  1.33 & 1.10 & 12.50 & 5.00   & 1.205  \\
Si~{\sc i}  & 10827.091   & ${}^3P^{o}_{2}$     & ${}^3P_{2}$ &  1.50  &  1.50 & 1.50 & 12.00 & 0.00   & 2.250  \\
He~{\sc i}  & 10830.25   & ${}^3S_{1}$     & ${}^3P^{o}_{1}$ &  2.00  &  1.50 & 1.75 & 4.00 & 0.00   & 2.875  \\
He~{\sc i}  & 10830.34   & ${}^3S_{1}$     & ${}^3P^{o}_{2}$ &  2.00  &  1.50 & 1.25 & 8.00 & -4.00   & 1.525  \\
	\hline
  \end{tabular}
  \egroup
\end{adjustbox}
  \vspace{0.2cm} 
 \label{index}   
\end{center}  
\end{table*}

We calculated the Land\'e factors for the \ion{Si}{i} line at 10827~\AA\ and the \ion{Ca}{ii} line at 8542~\AA\ by assuming LS coupling as in \cite{2004ASSL..307.....L}, that is,
\begin{subequations}
\begin{align}
g_{\rm eff} = & \frac{1}{2}(g_{1}+g_{2})+\frac{1}{4}(g_{1}-g_{2})[J_{1}(J_{1}+1)-J_{2}(J_{2}+1)], \\
G_{\rm eff} = & g^2_{\rm eff} - \frac{1}{80}\left(g_1-g_2\right)^2\left(16s-7d^2-4\right),
\end{align}
\end{subequations}
where $g_i$ and $J_i$ with $i=1,2$ are the Land\'e factor and total angular momentum of the upper and lower levels of the transition. The Land\'e factor and the quantities $s$ and $d$ are given by
\begin{subequations}
\begin{align}
g_i = & \frac{3}{2} + \frac{S_i(S_i+1)-L_i(L_i+1)}{2J_i(J_i+1)},\quad (i=1,2)\\
s = & J_1(J_1+1)+J_2(J_2+1), \label{seq} \\
d = & J_1(J_1+1)-J_2(J_2+1), \label{deq}
\end{align}
\end{subequations}
with $S_i$ and $L_i$ the spin and orbital angular momenta of the upper and lower levels of the transition. 

In Table~\ref{index}, we show, from left to right, the atomic species, line core wavelength, lower and upper level atomic configuration, Land\'{e} factor, effective Land\'{e} factor for the circular polarisation, $s$ and $d$ (see Eq.~\ref{seq} and \ref{deq}), and the effective Land\'{e} factor for the linear polarisation.

For the \ion{He}{i} triplet at 10830~\AA\, the situation is more complex. The red component of the multiplet, which is the one showing polarisation signals in our observations, is a blend of the two transitions listed in Table~\ref{index}. Consequently, for this spectral line, we calculated $g_{\rm eff}$ by applying the WFA to synthetic profiles obtained with the HAnle and ZEeman Light \citep[HAZEL,][]{2008ApJ...683..542A} code for several known magnetic fields and $\lambda_0=10830.34$~\AA. While the WFA fit is not perfect due to the blend, the best fit is achieved for $g_{\rm eff}=1.134$. Also, for the \ion{He}{i} line, we have only applied the WFA to the circular polarisation profiles, as the linear polarisation profiles cannot be modelled by considering only the Zeeman effect because there can be a significant contribution by the Hanle effect \citep{2007ApJ...655..642T}.

By assuming Gaussian and uncorrelated noise, we can write the maximum likelihood expressions for $B_\parallel$, $B_\perp$, and $\phi$ \citep{2012MNRAS.419..153M},
\begin{subequations}
\begin{align}
B_{\parallel} = & -\frac{1}{C\lambda_0^2 g_{eff}}\frac{\sum_\lambda V_\lambda \frac{\partial I(\lambda)}{\partial \lambda}}{\sum_\lambda \left( \frac{\partial I(\lambda)}{\partial \lambda} \right)^2}, \label{bpar} \\
B^2_{\perp} = & \frac{4}{\gamma C^2\lambda_0^4G_{eff}^2} \frac{\sqrt{\left(\sum_\lambda Q_\lambda \mathcal{I}\right)^2+\left(\sum_\lambda U_\lambda \mathcal{I}\right)^2}}{\sum_\lambda \mathcal{I}^2}, \label{bper}
\end{align}
\end{subequations}
where
\begin{equation}
\begin{cases}
\gamma=1, \\
\mathcal{I}=\frac{\partial^2 I(\lambda)}{\partial \lambda^2},
\end{cases}
\end{equation}
for the line centre, and
\begin{equation}
\begin{cases}
\gamma=3, \\
\mathcal{I}=\frac{1}{\lambda-\lambda_0}\frac{\partial I}{\partial \lambda},
\end{cases}
\end{equation}
for the line wing. We consider  for calcium as line wing and line core the ranges [8541.045, 8541.279] and [8541.881, 8542.349]~\AA, respectively. For the silicon and helium, we consider the range between [10825.050,10828.127] and [10830.017, 10831.644]~\AA, respectively.

The magnetic field inclination and azimuth in the plane of the sky are given by
\begin{subequations}
\begin{align}
\Theta = & \arctan \frac{B_{\perp}}{B_{\parallel}}, \label{binc} \\
\phi = & \frac{1}{2} \arctan \frac{\sum_\lambda U_\lambda}{\sum_\lambda Q_\lambda } + \phi_0, \label{bazi}
\end{align}
\end{subequations}
with the conditions for $\phi_0$ as defined in \cite{2012MNRAS.419..153M}.

\begin{figure}
	\begin{center} 
		\includegraphics[width=0.48\textwidth]{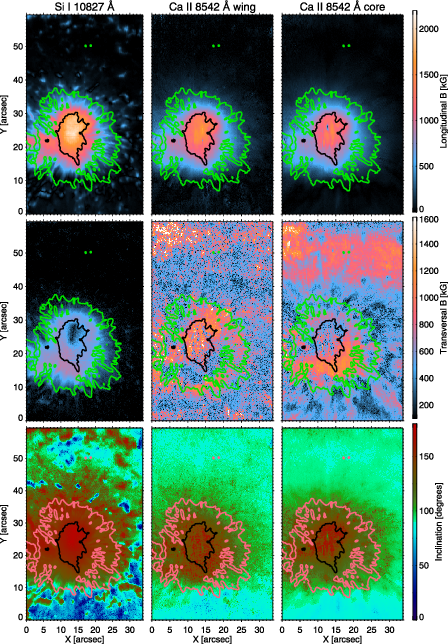}
		\caption{Spatial distribution of the longitudinal (top row) and transverse (middle row) components of the magnetic field inferred from the WFA on the \ion{Si}{i} line (first column) and on the \ion{Ca}{ii} line wing (second column) and line centre (third column). The bottom row shows the inclination of the magnetic field obtained by applying Eq.~\ref{binc} to the WFA results. The wavelength ranges considered in the calculations are indicated in the text. Contours tentatively define the area corresponding to the umbra and penumbra computed from the continuum intensity maps presented in Figure~\ref{Context}.}
		\label{Weak}
	\end{center}
\end{figure}

\begin{figure}
	\begin{center} 
		\includegraphics[width=.45\textwidth]{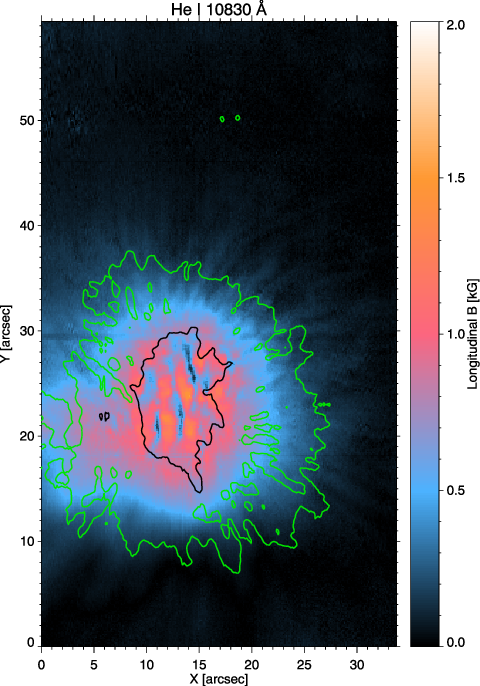}
		\caption{Spatial distribution of the longitudinal component of the magnetic field inferred from the application of the WFA to the \ion{He}{i} line. Contours tentatively define the area corresponding to the umbra and penumbra computed from the continuum intensity maps presented in Figure~\ref{Context}.}
		\label{Weak2}
	\end{center}
\end{figure}

In Fig.~\ref{Weak}, we present the parallel and perpendicular components of the magnetic field and the inclination with respect to the LOS (a LOS which is very close to the local vertical as we are close to the disc centre) resulting from the application of Eqs.~\eqref{bpar}, \eqref{bper}, and \eqref{binc} to the \ion{Si}{i} and \ion{Ca}{ii} lines in the aforementioned wavelength ranges. In Fig.~\ref{Weak2}, we show the parallel component of the magnetic field from the application of Eq.~\eqref{bpar} to the \ion{He}{i} line in the abovementioned wavelength range. Unlike the \ion{Ca}{ii} line, the \ion{Si}{i} and \ion{He}{i} lines are sensitive to the physical parameters of the solar atmosphere in a relatively narrow range of heights, and the inferred $B_\parallel$ does not show significant variation when restricting Eq.~\eqref{bpar} to different wavelength ranges. Thus, we only show one value for the longitudinal magnetic field for these spectral lines. Moreover, as mentioned above, the WFA is not suitable to estimate $B_\perp$ from the \ion{He}{i} line due to the potential impact of the Hanle effect on the linear polarisation profiles, particularly on regions where the magnetic field is weak. Given the height of formation of the observed \ion{Si}{i} and \ion{Ca}{ii} lines \citep[see, for instance,][respectively]{2024A&A...692A.169Q,QuinteroNoda2016}, moving from left to right in Fig.~\ref{Weak} is roughly equivalent to moving up in height in the solar atmosphere.

\begin{figure*}
	\begin{center} 
		\includegraphics[trim=0 0 0 0,width=18.0cm]{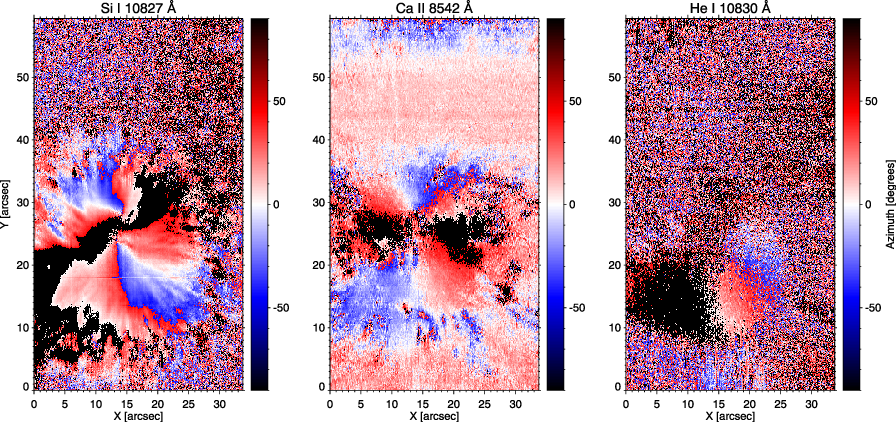}
		\caption{Spatial distribution of the magnetic field azimuth in the plane of the sky inferred with Eq.~\eqref{bazi} for the \ion{Si}{i} line at 10827~\AA\ (left), the \ion{Ca}{ii} line at 8542~\AA \ (middle), and the \ion{He}{i} line at 10830~\AA\ (right). The wavelength ranges considered in the calculation are indicated in the text, and we have used only the line centre range for the \ion{Ca}{ii} line for this plot.}
		\label{Azi}
	\end{center}
\end{figure*}

In the deeper layers, corresponding to the \ion{Si}{i} spectral line, we infer a strong $B_\parallel$ (top row in Fig.~\ref{Weak}) in the inner penumbra, in the umbra, and the penumbra to the left of the sunspot, with the rest of the penumbra showing weak $B_\parallel$. The rest of the FOV is dominated by weaker $B_\parallel$, with some scattered features corresponding to network concentrations. At the height sampled by the \ion{Ca}{ii} line wing, $B_\parallel$ has a similar behaviour. Finally, at the regions sampled by the \ion{Ca}{ii} line centre, there is a thread-like pattern in the umbra produced by the umbral flashes, where the inferred $B_\parallel$ slightly increases with respect to the values at wing wavelengths. This may be due to a change in the height of formation (e.g., the line is sensitive to lower layers in the atmosphere) instead of an actual increase in the magnetic field strength. The magnetic field inferred from the \ion{He}{i} line shows a similar spatial distribution (see Fig.~\ref{Weak2}) where the threaded-like pattern in the umbra due to the presence of umbral flashes becomes one of the most prominent features again.

Regarding the inferred $B_\perp$ (middle row in Fig.~\ref{Weak}), it is essential to emphasise that this quantity (and thus the inclination angle) has a non-zero bias and therefore small $B_\perp$ (at the order or smaller than the noise amplitude) are overestimated \citep{2012MNRAS.419..153M}. Especially for the \ion{Ca}{ii} line, the observed linear polarisation signals are below the noise level in a significant amount of pixels (see Fig.~\ref{Spatial_spectra}). At the height sampled by the \ion{Si}{i}, we infer relatively strong $B_\perp$ in most of the penumbra and weak values for the inner umbra. The \ion{Ca}{ii} line wing shows strong $B_\perp$ values in the complex light-bridge-like structure and the penumbra. In contrast, the results from the line centre show significant values in the lower (southern) part of the penumbra. Notably, many quiet Sun areas show a transverse magnetic field in the order of kG for the inferred transverse field for the \ion{Ca}{ii} line. We do not consider this result plausible for chromospheric quiet Sun magnetic fields, and we believe it is because most of the pixels outside the active region have linear polarisation signals that are close to or below the noise level.

Figure~\ref{Azi} displays the inferred magnetic field azimuth in the plane of the sky. We note that we have not disambiguated the azimuth; thus, the values are in the $[-90^\circ,90^\circ]$ range. For both the observed \ion{Si}{i} and \ion{Ca}{ii} lines, the inferred azimuth closely follows an expected radial pattern with respect to the centre of the sunspot. Although the values inferred from the \ion{Ca}{ii} line seem slightly shifted with respect to those corresponding to the \ion{Si}{i} line. This could be due to a rotation of the magnetic field with height. Still, more sophisticated inference techniques, which we plan to apply in a forthcoming work, are necessary to confirm this interpretation. Finally, the \ion{He}{i} transition only shows linear polarisation signals in the lower part of the penumbra, and the results in this region are only partially consistent with those corresponding to the \ion{Si}{i} and \ion{Ca}{ii} lines. On the one hand, the azimuth values for penumbral regions between $X=[15,23]$~arcsec seem consistent with those inferred for the other two transitions. On the other hand, values obtained outside the mentioned penumbral region are quite different than those from the other two transitions. Thus, if we consider that the linear polarisation signals for the helium transition are very weak almost everywhere (see Figure~\ref{Spatial_spectra}), we should not pay too much attention to the helium results.

\section{Summary}

We have presented the new capabilities of the upgraded GRIS instrument. The new two-channel configuration allows for the simultaneous observation of the \ion{Ca}{ii} line at 8542~\AA, the \ion{Si}{i} line at 10827~\AA, and the \ion{He}{i} triplet at 10830~\AA, probing a wide region of the solar atmosphere. In the future, a third spectral channel at 770~nm will be added, covering the \ion{K}{i} D$_1$ line, which probes a range of heights complementary to the currently available channels \citep[][]{2017MNRAS.470.1453Q}. In addition, while the observations shown in this work were acquired by scanning with the long-slit, the GRIS instrument is also equipped with an integral field unit based on image slicers, which will be offered to observers starting the first semester of 2025. This upgrade to GRIS will lead to new and exciting research.

In this work, we inferred some plasma properties, namely the longitudinal component of the velocity and the magnetic field vector, at several heights in the solar atmosphere by applying simple inference techniques to the observed spectra. Inferring the stratification of these quantities has been possible with the upgraded GRIS thanks to the simultaneous observation of spectral lines formed at different heights in the solar atmosphere. The results of the inference with these simple techniques are consistent with the current knowledge regarding the typical magnetic field vector in sunspots \citep[see, for instance, the reviews of][]{2003A&ARv..11..153S,2011LRSP....8....4B,2019LRSP...16....1B}, giving us confidence in the good performance of the instrument during the commissioning season.

However, properly exploiting these new data requires performing non-local thermodynamic equilibrium inversions of the \ion{Si}{i} and \ion{Ca}{ii} lines. Moreover, several photospheric atomic and molecular lines are present in the observed spectra and can be helpful to constrain better the inversions in the deepest layers of the photosphere. In a follow-up work, we plan to perform such inversions in the observed FOV, analysing in further detail the ensuing inferred stratification of physical quantities.

\begin{acknowledgements}
We thank the Gregor team, particularly the campaign's observing assistant, Jürgen Rendtel, for helping perform this observation and obtain such high-quality data. C. Quintero Noda, J. C. Trelles Arjona, T. del Pino Alem\'an, and M. J. Martínez González, acknowledge support from the Agencia Estatal de Investigación del Ministerio de Ciencia, Innovación y Universidades (MCIU/AEI) under grant ``Polarimetric Inference of Magnetic Fields'' and the European Regional Development Fund (ERDF) with reference PID2022-136563NB-I00/10.13039/501100011033. The publication is part of Project ICTS2022-007828, funded by MICIN and the European Union NextGenerationEU/RTRP. T. del Pino Alem\'an's participation in the publication is part of the Project RYC2021-034006-I, funded by MICIN/AEI/10.13039/501100011033, and the European Union ``NextGenerationEU''/RTRP. T. Felipe acknowledges grants PID2021-127487NB-I00, CNS2023-145233 and RYC2020-030307-I funded by MCIN/AEI/10.13039/501100011033.
\end{acknowledgements}

\bibliographystyle{aa} 
\bibliography{aa53432-24} 

\end{document}